\documentclass{aa}  
\usepackage{graphicx}
\usepackage{txfonts}
\begin{document}
   \title{On the chemistry and distribution of HOC$^+$ in M~82}
   \subtitle{More evidence for extensive PDRs}

   \author{A.~Fuente
          \inst{1}
          \and
          S.~Garc\'{\i}a-Burillo
          \inst{1}
          \and
          A.~Usero
          \inst{1,2}
          \and
          M.~Gerin
          \inst{3} 
          \and
          R.~Neri
          \inst{4}
         \and
          A.~Faure
         \inst{5}
          \and
          J.~Le~Bourlot
         \inst{6}
          \and
          M.~Gonz\'alez-Garc\'{\i}a
          \inst{6}
          \and
          J.R.~Rizzo
         \inst{7}
          \and
          T.~Alonso-Albi
          \inst{1}
          \and
          J.~Tennyson
	 \inst{8}
}

   \offprints{A. Fuente}

   \institute{ Observatorio Astron\'omico Nacional (OAN), Apdo. 112,
             E-28800 Alcal\'a de Henares, Madrid, Spain
	      \and
             Centre for Astrophysics Research, University of Hertfordshire, College
             Lane, Hatfield AL10 9AB, United Kingdom
             \and
             Laboratoire d'Etude du Rayonnement et de la Matiere, UMR 8112, CNRS, Ecole
             Normale Superieure et Observatoire de Paris, 24 rue Lhomond, F-75231 Paris Cedex 05, France
             \and
             Institut de Radioastronomie Millim\'etrique, 300 rue de la Piscine, F-38406 St Martin d'H\`eres Cedex, France
             \and
             Laboratoire d'Astrophysique Observatoire de Grenoble, BP 53, F-38041 Grenoble C\'edex 9, France
             \and
             LUTH, Observatoire de Paris and Universit\'e Paris, 7 Place Jansen, F-92190 Meudon, France
             \and
             Laboratorio de Astrof\'{\i}sica Espacial y F\'{\i}sica Fundamental, Apdo 78, E-28691 Villanueva de la Ca\~nada, Madrid, Spain
             \and
             Department of Physics and Astronomy, University College London, Gower Street, London WC1E 6BT, United Kingdom}
   \date{}

 
  \abstract
   {The molecular gas composition in the inner 1~kpc disk of the starburst galaxy M~82 resembles that of Galactic Photon
 Dominated Regions (PDRs). In particular, large abundances of the reactive ions HOC$^+$ and CO$^+$ have been measured in the nucleus of this galaxy. Two explanations have been proposed for such large abundances : the influence of intense UV fields from massive stars, or a significant role
of X-Rays.}
   {Our aim is to investigate the origin of the large abundances of reactive ions in M~82. }
   {We have completed our previous 30m HOC$^+$ J=1$\rightarrow$0 observations with the higher excitation HCO$^+$ and HOC$^+$ J=4$\rightarrow$3 and 3$\rightarrow$2 rotational lines. In addition, we have obtained with the IRAM Plateau de Bure Interferometer (PdBI) a 4$"$ resolution map of the HOC$^+$ emission in M~82, the first ever obtained in a Galactic or extragalactic source. }
   {Our HOC$^+$ interferometric image shows that the emission of the HOC$^+$ 1$\rightarrow$0 line is mainly restricted to the nuclear disk, with the maxima towards the E. and W. molecular peaks. In addition, line excitation calculations imply that the HOC$^+$ emission arises in dense gas (n$\ge$10$^4$~cm$^{-3}$). Therefore, the HOC$^+$ emission is arising in the dense PDRs embedded in the M~82 nuclear disk, rather than in the intercloud phase and/or wind.} 
   {We have improved our previous chemical model of M~82 by (i) using the new version of the Meudon PDR code, 
(ii) updating the chemical network, and (iii) considering two different types of clouds (with different thickness)
irradiated by the intense interstellar UV field (G$_0$=10$^4$ in units of the Habing field) prevailing in the nucleus of M~82. 
Most molecular observations (HCO$^+$, HOC$^+$, CO$^+$, CN, HCN, H$_3$O$^+$) are well explained assuming that $\sim$87\% of the 
mass of the molecular gas is forming small clouds (A$_v$=5~mag) while only $\sim$13\% of the mass is in large molecular clouds (A$_v$=50~mag). Such small number of large molecular clouds suggests that M~82 is an old starburst, where star formation has almost exhausted the molecular gas reservoir.}

   \maketitle
%

\section{Introduction}

M~82 is one of the nearest and brightest starburst galaxies. Located at a distance of 3.9~Mpc, 
and with a luminosity 
of 3.7$\times$10$^{10}$~L$_{\odot}$, it has been the subject of 
continuum and line observations in virtually all wavelengths from
X-rays to the radio domain.
Several molecular line studies indicate that the strong UV
field has heavily influenced the physical conditions, kinematics and chemistry in M~82
(Mao et al. 2000, Lord et al. 1996). 

On the other hand, M~82 exhibits one of the largest optically visible outflows or $``$superwinds" in the local
Universe. Superwinds are believed to be driven by the thermal and ram pressure of an initially very hot
(T$\sim$10$^8$~K), high pressure ($P/k$$\sim$10$^7$~K~cm$^{-3}$) and low density wind, itself
created from merged supernovae remnants, and to a lesser extent by the stellar winds from massive
stars. This very hot component is the responsible for the X-ray emission detected in this starburst galaxy
(Strickland \& Heckman 2007).

Back to the millimeter observations,
M~82 is one of the most beautiful examples of how chemistry can help to the full understanding
of the interstellar medium (ISM) of an external galaxy and the reverse, how the extragalactic research can
push forward the chemistry comprehension. Our early HCO interferometric map using the PdBI showed that
the M~82 nucleus is a giant photon-dominated region (PDR) of $\sim$650~pc size. 
Furthermore the comparison between the HCO and H$^{13}$CO$^+$ images suggested that 
the PDR chemistry is propagating across the M~82 nucleus (Garc\'{\i}a-Burillo et al. 2002, hereafter Paper I). 
This propagation has to be understood in terms of the existence of different populations
of clouds (different sizes, densities and temperatures) bathed by an intense UV field and with 
different spatial distribution across the M~82 disk. The densest clouds
seem to be concentrated in two $``$hot spots" located $\pm$15$"$ from the center of the galaxy, E. and W. peaks
(Mauersberger \& Henkel 1991).
Our subsequent chemical study using the 30m telescope allowed us to put
some restrictions to the ISM physical conditions. We observed the
well known PDR tracers CN, HCN, HOC$^+$, and CO$^+$ towards the M~82 nucleus 
(Fuente et al. 2005, 2006; hereafter Paper II and III). The high [CN]/[HCN] ratio
and the detection of the  reactive ions HOC$^+$ and CO$^+$ all across the galaxy disk
proves the exceptionally high energetic conditions prevailing in the ISM of
this galaxy. In extragalactic research, HOC$^+$  has only been detected in the active nucleus 
of NGC 1068 (Usero et al. 2004) and CO$^+$ has been only tentatively 
detected in Cygnus A (Fuente et al. 2000). 

In Paper II and III, we modeled the chemistry in the M~82 disk using a PDR chemistry with an
enhanced cosmic ray flux (4~10$^{-15}$~s$^{-1}$). Our best model accounted for the
observed [CN]/[HCN] and [CO$^+$]/[HCO$^+$] ratios in the scenario of a highly fragmented 
interstellar medium in which small dense cores (n$\sim$10$^5$~cm$^{-3}$, 
N$_{tot}$$<$1.3$\times$10$^{22}$~cm$^{-2}$ ) are bathed by an intense UV field (G$_0$=10$^4$ Habing
fields). This scenario is consistent with that proposed by Lord et al. (1996) based on observations
of FIR forbidden lines. Melo et al. (2005) catalogued a total of 197 young
massive clusters only in the starburst core. This 
incredible density of star clusters gives rise to a very unusual, highly energetic, and 
highly fragmented medium which is very different from anything in our Galaxy.
The main drawback of our model is that it was unable to reproduce the large CO$^+$
column densities measured in this galaxy.

PDR chemistry is not the unique proposed scenario to explain the molecular chemistry in
M~82. Spaans \& Meijrenik (2007) suggested that the high CO$^+$ and HOC$^+$
column densities could be explained if their emission arises in a X-ray dominated
region (XDR). Recently, van der Tak et al. (2008) detected the molecular 
ion H$_3$O$^+$ in M~82 and concluded that the observed 
H$_3$O$^+$ column densities were better explained by the enhanced cosmic 
ray PDR model than by the XDR model, however.

Our aim now is to investigate the origin, UV photon dominated or X-ray dominated chemistry, 
of the large abundance of reactive ions in M~82. 
In this Paper, we present a 4$"$ angular resolution image of
the HOC$^+$ 1$\rightarrow$0 line towards M~82, the first
ever obtained in a Galactic or extragalactic source. In addition, we determine the excitation
conditions of the gas emitting in HCO$^+$ and HOC$^+$. Our results discard X-rays as the major driver
of the reactive ions chemistry in M~82. With this information in our hands, we
improve our chemical PDR model to account for the molecular abundances measured
in M~82 thus far.

M~82 is one of the few galaxies in which we can resolve the individual molecular clouds
and provides an excellent opportunity to study the PDR and/or XDR chemistries within a single
object. This kind of studies are of paramount importance to establish extragalactic chemical patterns to
discern between the influence of soft X-rays and UV photons on the chemistry of the
most distant galaxies.

\vskip 0.5cm
\setlength\unitlength{1cm}
\begin{figure}
\vspace{5cm}
\includegraphics{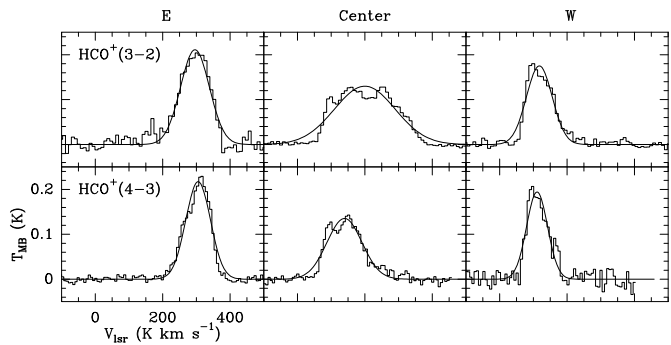}
\caption{Spectra of the J=3$\rightarrow$2 and J=4$\rightarrow$3 rotational lines of HCO$^+$ as observed
with the JCMT. The W., Center and E. positions were observed in the two lines. Intensity scale is 
main beam temperature. } 
\label{fig1}
\end{figure}

\vskip 0.5cm
\setlength\unitlength{1cm}
\begin{figure*}
\vspace{13cm}
\includegraphics{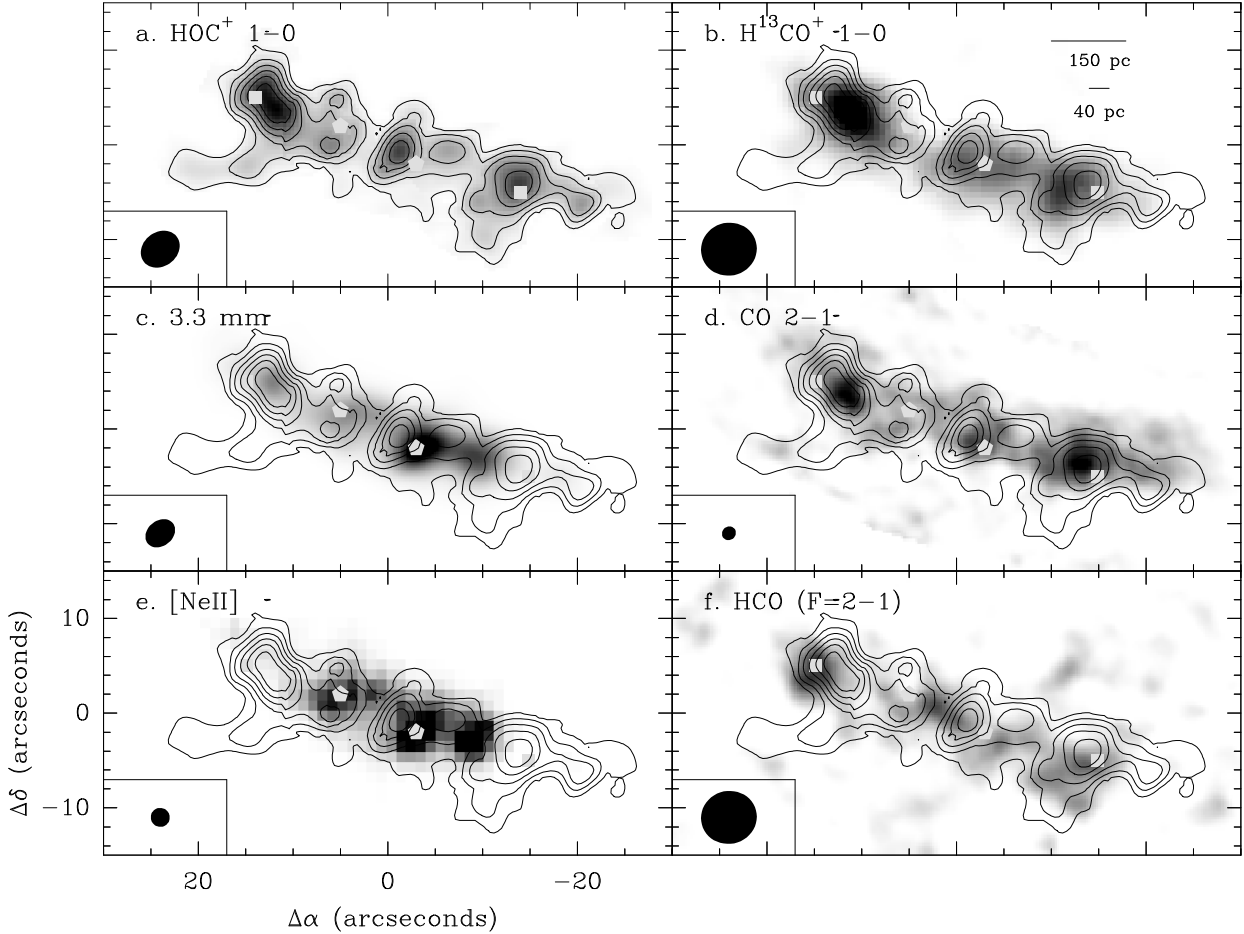}
\caption{{\bf a.} Integrated intensity image of the HOC$^+$ 1$\rightarrow$0 line as observed
with the PdB interferometer. Contours are 31.8, 77.7, 124.3, 170.9, 217.6 and 264.2 mJy/beam $\times$ km/s. 
The inner ring is indicated by white pentagons and 
the outer ring by white squares. {\bf b.} The integrated intensity map of the HOC$^+$ 1$\rightarrow$0 
line superposed on the interferometric H$^{13}$CO$^+$ 1$\rightarrow$0 image reported in 
Paper I (grey scale: 100--800 mJy/beam $\times$ km/s ). {\bf c.} 
The same as {\bf b.} but on our continuum interferometric image at 3.3mm (grey scale: 1--35 mJy/beam). 
{\bf d.} The same as {\bf b.} but on the interferometric continuum $^{12}$CO 2$\rightarrow$1 image 
reported by Weiss et al. (2001) (grey scale: 0.7--2.5 Jy/beam $\times$ km/s). {\bf e.} 
The same as {\bf b.} but on the [NeII] image published by 
Achertman \& Lacy (1995) (grey scale: 0.110--0.90 Jy). 
{\bf f.} The same as {\bf b.} but on the HCO image published in 
Paper I (grey scale: 100--350 mJy/beam $\times$ km/s).} 
\label{fig1}
\end{figure*}

\section{Observations}

\subsection{Single-dish observations}
Observations of the J=3$\rightarrow$2 and J=4$\rightarrow$3 rotational lines of HCO$^+$
and HOC$^+$ were carried out using the 30m IRAM telescope located at Pico de Veleta (Spain) and
the JCMT telescope at the Mauna Kea (Hawaii). In Table 1 we show a summary of all the 
observations used in this Paper. The 30m telescope observations were presented in Paper III. 

In Fig. 1 and Table 2 we show
the observational results from the JCMT observations. 
All the lines were observed with a spectral resolution of 1~MHz. The frequencies of
the rotational lines of both isotopomers, HCO$^+$ and HOC$^+$,
differ by less than 1~GHz. Therefore,
we observed them simultaneously avoiding possible pointing and
calibration errors in the measurement of the relative intensities.

\subsection{Interferometric observations}
We present high-angular resolution observations in the continuum
at 3.3~mm and in the HOC$^+$ 1$\rightarrow$0 rotational lines towards M~82.
The observations were carried out with the IRAM\footnote{IRAM is 
supported by INSU/CNRS (France), MPG (Germany) and IGN (Spain). } array at 
Plateau de Bure (PdBI) in 2005, May and December. The antennas were arranged in
the C and D configurations providing an almost circular beam of 4.4$"$ $\times$ 3.6$"$ 
(natural weighting). We adjusted the spectral correlator to give a contiguous
bandwidth of 1~GHz. The frequency resolution was set
to 2.5~MHz (8.4~km/s).  We subtracted the 3.3mm continuum emission 
using the GILDAS software package to generate the 
HOC$^+$ 1$\rightarrow$0 image. The noise (rms) in the spectroscopic 
image is of $\sim$1~mJy/beam.

In addition, a 3.3 mm continuum map was generated by averaging the channels
free of line emission. Uniform weighting was applied to the
measured visibilities producing a clean beam of 3.5$"$ $\times$ 2.7$"$. 
The noise (rms) in the 3.3mm continuum image is 0.3 mJy/beam.

The maps have not been corrected for primary beam attenuation which
has at the frequency of HOC$^+$ a HPBW of 56$"$ . The images are
centered at the dynamical center of the galaxy,
R.A.= 09:55:51.9, Dec=+69:40:46.90 (J2000).

\begin{table}
\begin{center}
\caption{Line and telescope parameters\label{tbl-1}}
\begin{tabular}{llcccc}
\hline\hline
Molecule & Transition & Freq. (GHz) & Tel. & HPBW \\
\hline
H$^{13}$CO$^+$  &  J=1$\rightarrow$0         & 86.75433     & PdBI &  5.9$"$$\times$5.6$"$  \\ 
HOC$^+$         &  J=1$\rightarrow$0         & 89.48741     & PdBI &  4.38$"$$\times$3.6$"$ \\ \hline 
HOC$^+$         &  J=1$\rightarrow$0         & 89.48741     & 30m  &  28$''$ \\
HCO$^+$         &  J=3$\rightarrow$2         & 267.55753    & 30m  &  9$"$\\
HOC$^+$         &  J=3$\rightarrow$2         & 268.45109    & 30m  &  9$"$\\ \hline
HCO$^+$         &  J=3$\rightarrow$2         & 267.55753    & JCMT &  18$"$\\
HOC$^+$         &  J=3$\rightarrow$2         & 268.45109    & JCMT &  18$"$\\ 
HCO$^+$         &  J=4$\rightarrow$3         & 356.73413    & JCMT &  14$"$\\
HOC$^+$         &  J=4$\rightarrow$3         & 357.92199    & JCMT &  14$"$\\ 
\hline
\end{tabular}
\end{center}
\end{table} 

\begin{table}
\caption{Gaussian fits to the lines observed with the JCMT}
\begin{tabular}{lrcccc}
\hline\hline
 \multicolumn{2}{l}{}               & $\int T_{MB} dv$ &  $v_{lsr}$          &   $\Delta v$        & T$_{MB}$  \\ 
\multicolumn{2}{l}{Transition}      &  (K km s$^{-1}$)  &  (km s$^{-1}$)  &   (km s$^{-1}$)  & (mK)              \\ \hline
\multicolumn{6}{c}{E. (+14$"$,+5$"$)}                                                                                                            \\ \hline
HCO$^+$               &  3$\rightarrow$2   & 22.6 (0.2)    &    296(2)  &  101(3)   & 210(11) \\
HCO$^+$               &  4$\rightarrow$3   & 19.9 (0.3)    &    305(1)  &  86(2)    & 217( 7) \\
HOC$^+$               &  3$\rightarrow$2   & $<$1$^b$          &            &  101$^a$     & $<$33$^b$ \\
HOC$^+$               &  4$\rightarrow$3   & $<$0.5        &            &  86$^a$      & $<$21 \\
\hline 
\multicolumn{6}{c}{Center (0$"$,0$"$)}                                                                                                  \\ \hline
HCO$^+$               &  3$\rightarrow$2  & 29.4(0.2)     &    200(1)  &  212(2)      &  130(3) \\
HCO$^+$               &  4$\rightarrow$3  & 17.6(0.3)     &    140(2)  &  122(3)      &  135(8) \\
HOC$^+$               &  3$\rightarrow$2  & $<$0.4        &            &  212$^a$     & $<$9 \\
HOC$^+$               &  4$\rightarrow$3  & $<$0.7        &            &  122$^a$     & $<$24 \\
\hline
\multicolumn{6}{c}{W. (-14$"$,-5$"$)}                                 \\ \hline
HCO$^+$               &  3$\rightarrow$2 &  16.8(0.3)     &    118(1)   &  90(2)       & 174(7)  \\
HCO$^+$               &  4$\rightarrow$3  & 15.1(0.5)     &    111(1)   &  73(3)       & 194(17)  \\
HOC$^+$               &  3$\rightarrow$2  &  $<$0.6       &             &  90$^a$      & $<$21 \\
HOC$^+$               &  4$\rightarrow$3  &  $<$1.2       &             &  73$^a$      & $<$51 \\
\hline
\end{tabular}

\noindent
$^a$ Assumed linewidth to derive the integrated intensity upper limit.
 
\noindent
$^b$ Upper limits are 3$\sigma$.
\end{table}

\section{Results}

\subsection{Single-dish observations}
The HCO$^+$ 3$\rightarrow$2 and 4$\rightarrow$3 lines have been detected
towards the three positions observed across the M~82 disk (see Fig.~1).
We have not detected the HOC$^+$ 3$\rightarrow$2 and 4$\rightarrow$3 lines
towards any of the observed positions using the JCMT. Gaussian fits to the
HCO$^+$ 3$\rightarrow$2 and 4$\rightarrow$3 lines and upper limits to the
HOC$^+$ lines are given in Table 2.

To estimate the physical conditions of the emitting gas, mainly the molecular hydrogen density,
it is essential to derive accurate values of the (HCO$^+$ 4$\rightarrow$3)/(HCO$^+$ 3$\rightarrow$2) 
line intensity (T$_{B}$) ratio (see Sect. 4 and Fig. 4). Since the two lines are observed with different beams, 
this requires of making some assumptions about the emission
spatial distribution. On basis of our H$^{13}$CO$^+$ 1$\rightarrow$0 and 
HOC$^+$ 1$\rightarrow$0 interferometric images we have assumed that 
the emission of the HCO$^+$ 3$\rightarrow$2 and 4$\rightarrow$3 lines arises in a uniform
slab which is unresolved in the direction perpendicular to the plane. In this case we only need to 
make a 1-D beam dilution correction to the intensity of the J=4$\rightarrow$3 line to get 
the right line intensity  ratio. {\it With the assumed geometry, the  
HCO$^+$ 4$\rightarrow$3/HCO$^+$ 3$\rightarrow$2 line intensity 
ratio is $\sim$0.8 all across the galaxy plane.} 
We next discuss the quality of our approximation by comparing 
the intensities of the HCO$^+$ 3$\rightarrow$2 line as measured 
with the 30m telescope (HPBW$\sim$9$"$) 
with those measured with the JCMT (HPBW$\sim$18$"$).

The linewidths of the HCO$^+$ 3$\rightarrow$2 line as observed
using the 30m and the JCMT telescopes are quite similar towards the E. 
With our assumption that the emitting region is a uniform slab, the ratio 
between the line intensities should 
be $\sim$2. The measured value is 1.4.
Taking into account possible relative calibration errors (30$\%$) between the two 
instruments, our measurements are quite compatible with the assumed geometry.

The linewidths of the HCO$^+$ 3$\rightarrow$2 lines observed
with the 30m and the JCMT are, however, very different towards the Center.
This is not surprising taking into account that the size of the JCMT beam
is twice that of the 30m telescope (see Table~1). 
Three different velocity components (C1 centered at 255~km/s,
C2 at 160~km/s and C3 at 95 km/s) appear in the profile of 
the HCO$^+$ 3$\rightarrow$2 line as 
observed with the JCMT (see Fig. 1). 
Only the component at 160~km/s appears in the 30m spectrum (see Paper III). 
Assuming again that the emission is uniform in a slab centered in
the disk plane and unresolved in the perpendicular direction, one would 
expect that the intensity 
measured with the JCMT is a factor of 2 lower than that measured with the 30m because 
of the different beams. The line intensity of the C2 component is indeed a factor $\sim$1.7 
lower than the main brightness temperature measured with the IRAM 30m telescope. 
However this rule does not apply to the integrated intensities because of the
contribution of the C1 and C3 components in the JCMT spectrum.

Therefore, we conclude that the assumed geometry is consistent with our
30m and JCMT observations. 

\subsection{Interferometric observations}
In Fig. 2 we compare our HOC$^+$ 1$\rightarrow$0 image with the high angular resolution images of M~82 in several 
molecular lines. Our continuum image at 3.3~mm and the [NeII] map by Achertman \& Lacy (1995) are also shown. 

All molecular maps depict the typical nested rings morphology, 
with two maxima located almost symmetrically on both sides of the
dynamical center of the galaxy. However, the angular distance
between both peaks is different from one species to another. This
suggests a chemical stratification of the ring structure.
The  [Ne II] emission (Achertman \& Lacy 1995) and the radio 
recombination lines (Rodr\'{\i}guez-Rico et al. 2004) define a ring of HII regions 
with a width of $\sim$10$"$-15$"$. The peaks of this ring are marked with pentagons in Fig. 1
and hereafter we will refer to it as the {\it inner ring}. The CO and H$^{13}$CO$^+$ maxima lies beyond the HII regions
defining a ring with a width of $\sim$25$"$--28$"$. The HCO emission extends further out in the disk with
a separation of $\sim$32$"$ between the peaks. The peaks of this ring are marked with squares in Fig. 1
and hereafter we will refer to it as {\it outer ring}. We interpreted this chemical stratification 
as the consequence of the propagation of the PDR chemistry outwards the disk (Paper I).

The emission of the HOC$^+$ 1$\rightarrow$0 line is weak in the {\it inner ring}. 
The HOC$^+$ emission is concentrated in two maxima located at the inner edge of the {\it outer ring}, 
at the same distance from the galaxy center as the peaks of the H$^{13}$CO$^+$ 1$\rightarrow$0 and 
CO 2$\rightarrow$1 lines. In fact, the HOC$^+$ 1$\rightarrow$0, 
H$^{13}$CO$^+$ 1$\rightarrow$0 and CO 2$\rightarrow$1 lines have similar spatial distribution
at scales of $\sim$~150~pc. This similarity breaks down 
at smaller scales, however. The E. maximum of HOC$^+$ is displaced $\sim$40~pc towards the North 
relative to that of H$^{13}$CO$^+$. Similar displacements ($\sim$40~pc) between the HOC$^+$ and H$^{13}$CO$^+$
maxima are observed at several positions in the western-half of the M~82 disk. 
The spatial distributions of the CO 2$\rightarrow$1 and HOC$^+$ 1$\rightarrow$0 
lines are quite different in the region enclosed by the {\it inner ring}. This is very likely
due to the different critical densities of the two transitions. Mao et al. (2000) claimed that the
bulk of the CO emission is arising in a warm low density interclump medium instead of in the dense clouds. 
The spatial distribution of the  HOC$^+$ 1$\rightarrow$0 line is certainly very different from that of
the HCO F=2$\rightarrow$1 line. The HOC$^+$ emission is displaced, in z-direction, towards
the North and, in radial direction, towards the center of the galaxy relative to that of HCO. 
A very different spatial distribution between HCO and the other PDR tracers H$^{13}$CO$^+$, HOC$^+$, 
and CO$^+$, is also observed in the Galactic
PDR associated to the ultracompact HII region Mon~R2 (Rizzo et al. 2003, 2005). 

\subsubsection{H$^{13}$CO$^+$/HOC$^+$}
To quantify the observed differences between the spatial distribution of the 
H$^{13}$CO$^+$~1$\rightarrow$0 and HOC$^+$~1$\rightarrow$0 lines,
we have calculated the map of the H$^{13}$CO$^+$~1$\rightarrow$0/HOC$^+$~1$\rightarrow$0 
brightness temperature ratio (hereafter, $R$) using our PdBI images convolved to 
the same angular resolution (5.9$"$$\times$5.6$"$).  
The resulting image is shown in Fig. 2.
$R$ takes values between 1 and 2 in most of the galaxy disk without 
any clear trend with the distance from the galaxy center. 
This is consistent with the uniform value of $R$ measured with
the 30m telescope (Paper II). Larger values, R$\ge$3, are found
in the southern half of the bubble associated
with the supernova remnant SNR~41.95+57.5, which produces a North-South gradient
in the value of $R$ towards the western-half of the M~82 disk. Since the excitation conditions of the 
H$^{13}$CO$^+$ 1$\rightarrow$0 and HOC$^+$ 1$\rightarrow$0 lines are quite similar, 
this displacement is very likely due to a gradient in the [H$^{13}$CO$^+$]/[HOC$^+$] 
abundance ratio. 
Assuming a $^{12}$CO/$^{13}$CO ratio of 89 and the two isotopomers arising in the
same phase, values of R between  $\sim$1 and $\sim$3 corresponds to
[HCO$^+$]/[HOC$^+$] between $\sim$44 and $\sim$136. These values are similar to
those found in Galactic PDRs (Fuente et al. 2003, Rizzo et al. 2003, Savage \& Ziurys 2004) 
and towards the AGN in NGC~1068 (Usero et al. 2004)
but more than one order of magnitude larger than those found in Galactic dense clouds 
where [HCO$^+$]/[HOC$^+$]=2000--6000 (Apponi \& Ziurys 1997).  

\setlength\unitlength{1cm}
\begin{figure}
\vspace{8cm}
\includegraphics{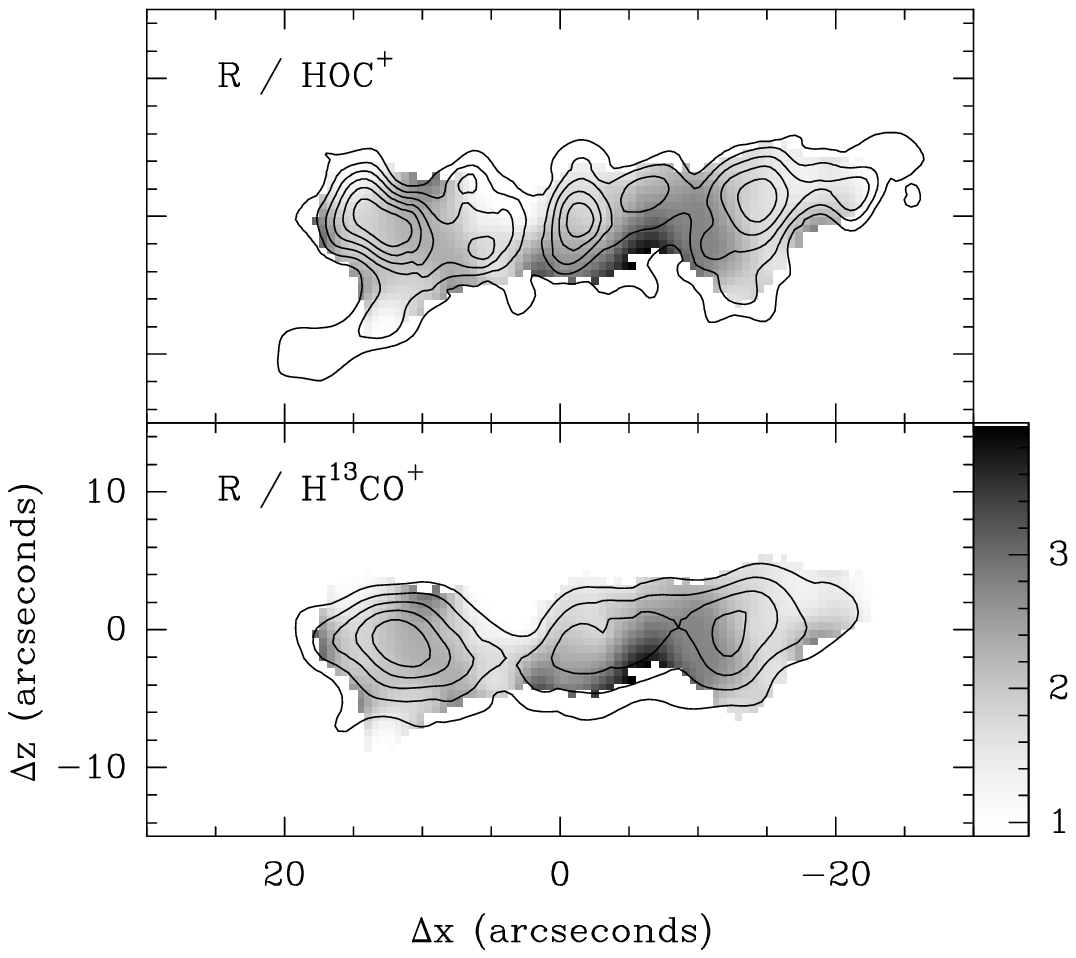}
\caption{Map of the H$^{13}$CO$^+$ 1$\rightarrow$0/HOC$^+$ 1$\rightarrow$0 brightness temperature
ratio, R, (grey scale) in M~82. The galaxy has been rotated to make coincide its major axis with the
abcissas axis. In the top panel we compare the H$^{13}$CO$^+$ 1$\rightarrow$0/HOC$^+$ 1$\rightarrow$0
image with that of the integrated intensity emission of the HOC$^+$ 1$\rightarrow$0 line, in the bottom,
with that of H$^{13}$CO$^+$ 1$\rightarrow$0 line (Paper~I).}
\label{fig2}
\end{figure}
\vskip 0.5cm

\setlength\unitlength{1cm}
\begin{figure*}
\vspace{10cm}
\includegraphics{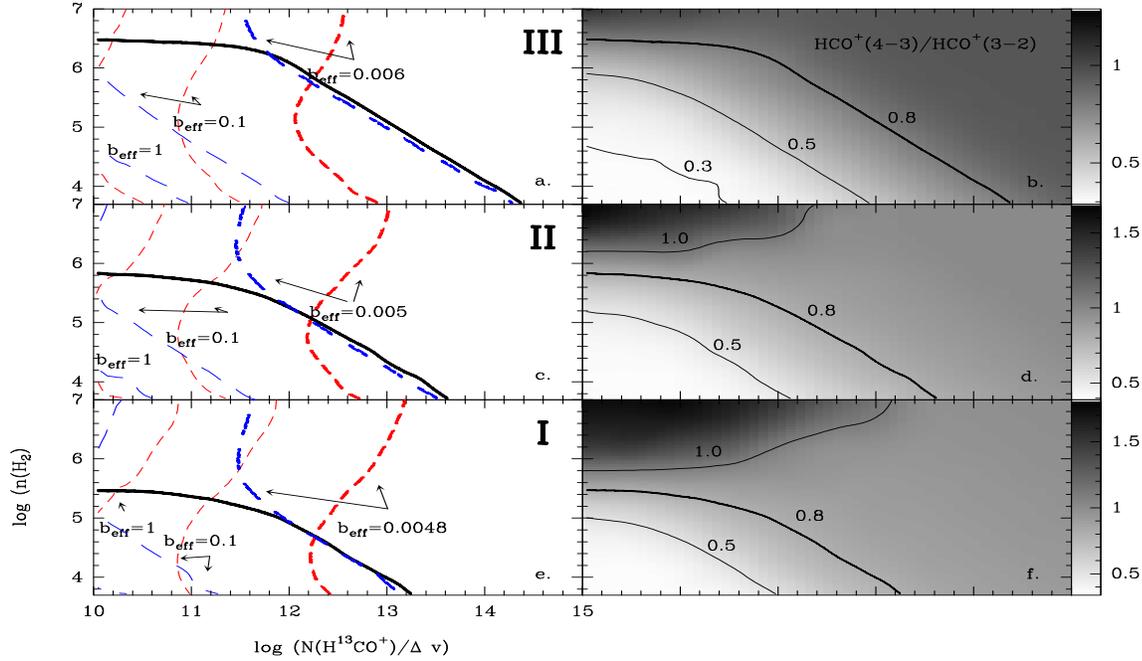}
\caption{Results from our LVG calculations assuming typical physical
conditions of the regions III ({\bf top}), II ({\bf middle}) and I ({\bf bottom}) 
of the PDR (see Fig. 6 and text).  In the right panels ({\bf b.},{\bf d.} and {\bf f.}) we show the
HCO$^+$ 4$\rightarrow$3 /HCO$^+$ 3$\rightarrow$2 line ratio as a function of 
the hydrogen density (n(H$_2$)) and the HCO$^+$ column density, 
N(HCO$^+$)= 89 $\times$ N(H$^{13}$CO$^+$). 
Left panels ({\bf a.}, {\bf c.} and {\bf e.}) have to be interpreted as follows:
the continuous line is the contour corresponding to
a HCO$^+$ 4$\rightarrow$3 /HCO$^+$ 3$\rightarrow$2 line intensity ratio of 0.8.
The HCO$^+$ 4$\rightarrow$3 /HCO$^+$ 3$\rightarrow$2 line intensity ratio does not
depend on the beam filling factor because we assume that it is the same for 
the two lines. Short-dashed lines correspond to an observed 
H$^{13}$CO$^+$ 1$\rightarrow$0 intensity of 9.1~mK (value observed
towards the E.) assuming different values of the beam filling factor for this line
emission. Long-dashed lines are the contours for a HCO$^+$ 3$\rightarrow$2 
intensity of 210~mK (value toward the E.) 
assuming a [HCO$^+$]/[H$^{13}$CO$^+$]=89 and the same values of the beam 
filling factor as for the H$^{13}$CO$^+$ line. Note that we only find a solution for a beam filling 
factor of 0.006 in region III, 0.005 assuming in region II and of 0.0048 
in region I. These are the solutions given in Table 3.}
\label{fig2}
\end{figure*}

\vskip 0.5cm
 
\section{LVG calculations}
We have made LVG calculations to derive the physical conditions of the emitting gas.
Chemical models show that high electron abundances, X(e$^-$)$>$10$^{-5}$, 
are required to have such low values of the [HCO$^+$]/[HOC$^+$] ratio, $<$80,
as measured in M~82 (see Paper II, Usero et al. 2004, Fuente et al.
2003). Therefore, part of the HCO$^+$ and most HOC$^+$ emission is very likely arising in the
partially ionized cloud envelopes and/or intercloud medium, and collisions with electrons
should be considered in our molecular excitation calculations.
We have calculated the HCO$^+$ and HOC$^+$ collisional rates with electrons using
the method described in Appendix A. While collisional rates with molecular hydrogen
are of the order of 10$^{-10}$ cm$^{-3}$ s$^{-1}$, the collisional rates with electrons are 
$\sim$~10$^{-5}$ cm$^{-3}$ s$^{-1}$, which corroborates that collisions with electrons 
are to be considered in our case (in which X(e$^-$)$>$10$^{-5}$).

We have derived the physical conditions of the gas emitting in HCO$^+$ by fitting the
spectra derived from the H$^{13}$CO$^+$ image smoothed to a beam of 18$"$,
and the JCMT observations.
In our LVG analysis we have assumed a constant [HCO$^+$]/{H$^{13}$CO$^+$] isotopic ratio of 89 and 
considered three different sets of kinetic temperature/electron density values that defined 
the typical stratification within a cloud irradiated by an intense UV field:  
(I) T$_k$=1000~K, X(e$^-$)=10$^{-3}$;
(II) T$_k$=200~K, X(e$^-$)=10$^{-4}$; and (III) T$_k$=50~K, X(e$^-$)=0.
These three regions are shown in Fig. 6 for a plane-parallel slab 
with a total visual extinction of 50~mag
with the two sides illuminated by the intense UV field prevailing in M~82.

In regions II and III the hydrogen is mainly in molecular form and collisions with
H$_2$ dominates the excitation of the HCO$^+$ molecules. However,
more than 50\% (the exact value depends on the exact value of the
visual extinction) of the hydrogen
is in atomic form in region I. Collisions with H could dominate in this region
because the rates for H  are usually larger than those for H$_2$ 
(partly because H is lighter than  H$_2$). As the rates for H are not known 
for this ion, we can consider that the present para-H$_2$ rates (Flower, 1999) can be used for 
H atoms within a  factor of 5. Such difference between 
H and H$_2$ collisions rates are indeed observed for CO 
(Flower 2001, Balakrishnan et al, 2002).
Thus our final estimate of the molecular hydrogen density in
region I is actually a particle (H and H$_2$) density and is accurate
within a factor of 5.

The results of our excitation calculations are shown in Table~3.
Since we have observed two lines of the main isotope HCO$^+$, and one of the rarer
isotope, H$^{13}$CO$^+$, we can determine the opacity of the lines and derive 
the beam filling factor, the H$^{13}$CO$^+$ 
(and therefore HCO$^+$) column density and the molecular hydrogen density 
from our LVG fitting.
Our observations are well accounted for by beam filling factors 
between 0.003-0.006, which at the distance of M~82 correspond to typical 
sizes of 20--30~pc, similar to the size of a giant molecular cloud in our Galaxy. 
Note, however, that this is an area beam filling factor and the linear size derived in this
way is only an upper limit to the actual cloud sizes.
The H$^{13}$CO$^+$ column density is N(H$^{13}$CO$^+$)/$\Delta$v$\sim$~2~10$^{12}$~cm$^{-2}$
towards the three positions. This value is quite robust since it is not strongly dependent 
on the assumed values of the kinetic temperature and electron abundance (see Fig.~4).
The derived values of n(H$_2$) do, however, depend on the assumed kinetic temperature and
electron abundance. 
For negligible values of X(e$^-$) we obtain molecular hydrogen densities larger
than 5$\times$10$^{5}$~cm$^{-3}$. The derived value of the 
molecular hydrogen density is lower by an order of magnitude if a kinetic temperature 
of 1000~K and an electron abundance of 10$^{-3}$ are assumed (see Table 3 and Fig. 4). 
The derived molecular hydrogen densities are larger than 
10$^4$~cm$^{-3}$ even in the case of the extreme physical conditions of region I. 
This proves that the bulk of the HCO$^+$ emission is arising from dense gas.

\begin{table}
\begin{center}
\caption{LVG results}
\begin{tabular}{llrccc}
\hline\hline
                  &   X(e$^-$)     &  b$_{eff}$     &  N(H$^{13}$CO$^+$)/$\Delta$v &  n(H$_2$)    &  T$_k$ \\ 
                  &                &                &  (cm$^{-2}$ km$^{-1}$ s)     &  (cm$^{-3}$) &  (K)        \\ \hline
E.                &    0           &  0.006         &  1.9 10$^{12}$               &  6.0 10$^5$  &   50        \\
Center            &    0           &  0.0035        &  1.5 10$^{12}$               &  7.4 10$^5$  &   50        \\
W.                &    0           &  0.0048        &  2.3 10$^{12}$               &  5.5 10$^5$  &   50        \\ \hline
E.                &    10$^{-4}$   &  0.005         &  1.7 10$^{12}$               &  1.1 10$^5$  &   200        \\
Center            &    10$^{-4}$   &  0.003         &  1.2 10$^{12}$               &  1.5 10$^5$  &   200        \\
W.                &    10$^{-4}$   &  0.004         &  2.2 10$^{12}$               &  9.2 10$^4$  &   200        \\ \hline
E.                &    10$^{-3}$    &  0.0048        &  1.7 10$^{12}$               &  5.0 10$^4$  &   1000        \\
Center            &    10$^{-3}$    &  0.0042        &  1.3 10$^{12}$               &  6.8 10$^4$  &   1000        \\
W.                &    10$^{-3}$    &  0.006         &  2.3 10$^{12}$               &  4.0 10$^4$  &   1000        \\ \hline
\hline
\end{tabular}
\end{center}
\end{table}

For HOC$^+$ our information is not complete. We have detected 
the HOC$^+$ 3$\rightarrow$2 line towards the E. using the 30m telescope 
(Paper III). We have not detected the HOC$^+$ 3$\rightarrow$2 line towards 
any position using the JCMT. Using the 30m data and our interferometric 
HOC$^+$ J=1$\rightarrow$0 image degraded to an angular resolution of 9$"$, 
we have made some LVG calculations and estimated the physical conditions of 
the gas emitting in HOC$^+$. Assuming T$_k$=50~K and X(e$^-$)=0, we derive a 
beam averaged HOC$^+$ column density , N(HOC$^+$)/$\Delta$v=2.8~10$^{12}$~cm$^{-2}$, 
and a molecular hydrogen density, n(H$_2$)$\sim$1~10$^5$~cm$^{-3}$, towards the E. 
The derived molecular hydrogen density is only a few times lower than that derived 
from the HCO$^+$ lines assuming the same physical conditions. As discussed in 
Section 5, these physical conditions  are not realistic for HOC$^+$ that is 
expected to mainly arise in region I. Assuming 
the most extreme physical conditions of region I, T$_k$=1000~K and X(e$^-$)=10$^{-3}$, 
the derived H$_2$ density is $\sim$ 2~10$^4$~cm$^{-3}$. 
This large density ($>$10$^4$~cm$^{-3}$) suggests that the HOC$^+$ emission arises 
in the partially ionized 
cloud envelopes rather than in the diffuse intercloud medium proposed by 
Mao et al. (2000) to explain the CO observations. 

\begin{figure*}
 \includegraphics[viewport=0 0 500 580, width=0.5\textwidth, height=0.5\textwidth, angle=0]{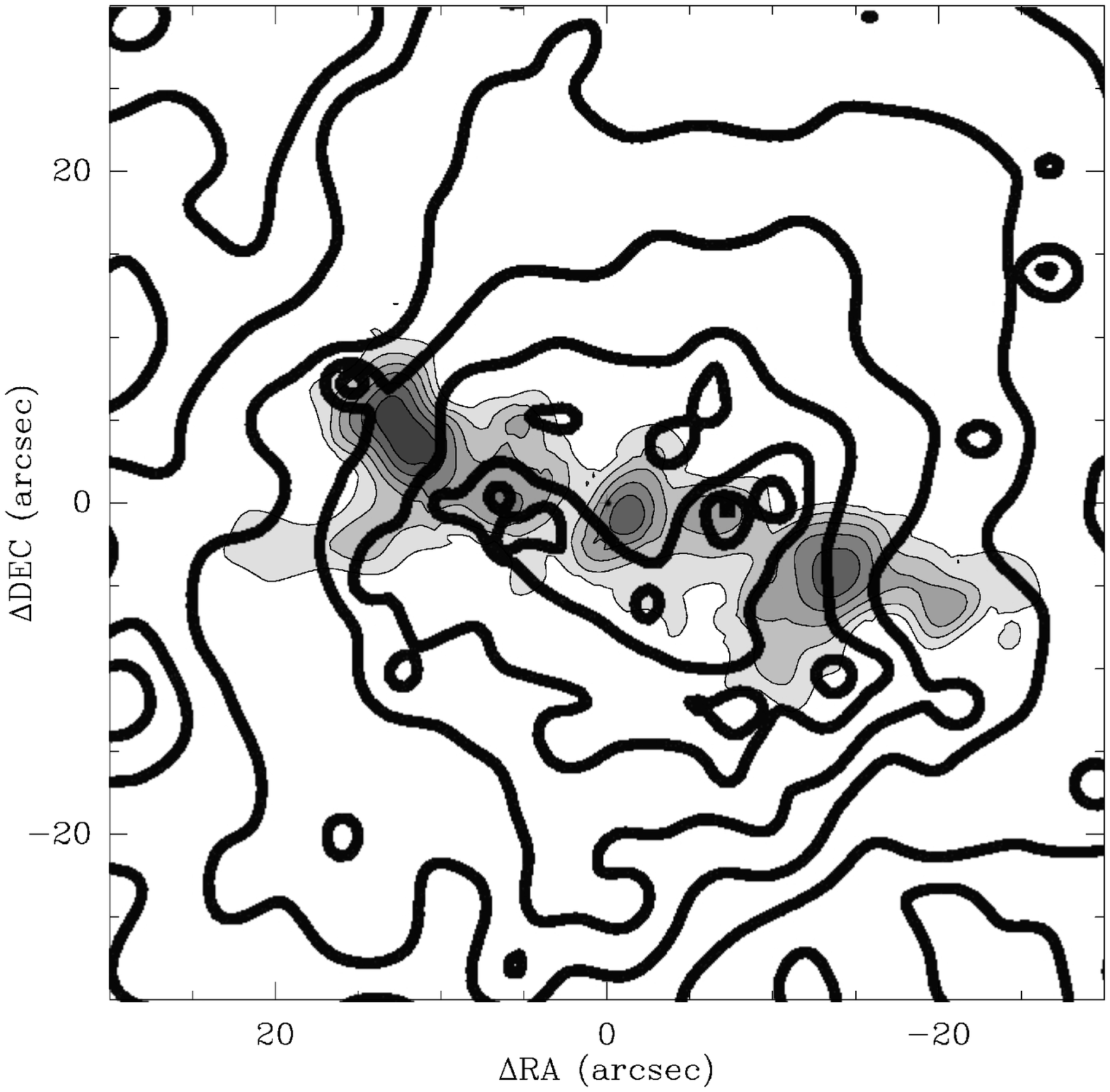}
 \includegraphics[viewport=0 0 500 580, width=0.5\textwidth, height=0.5\textwidth, angle=0]{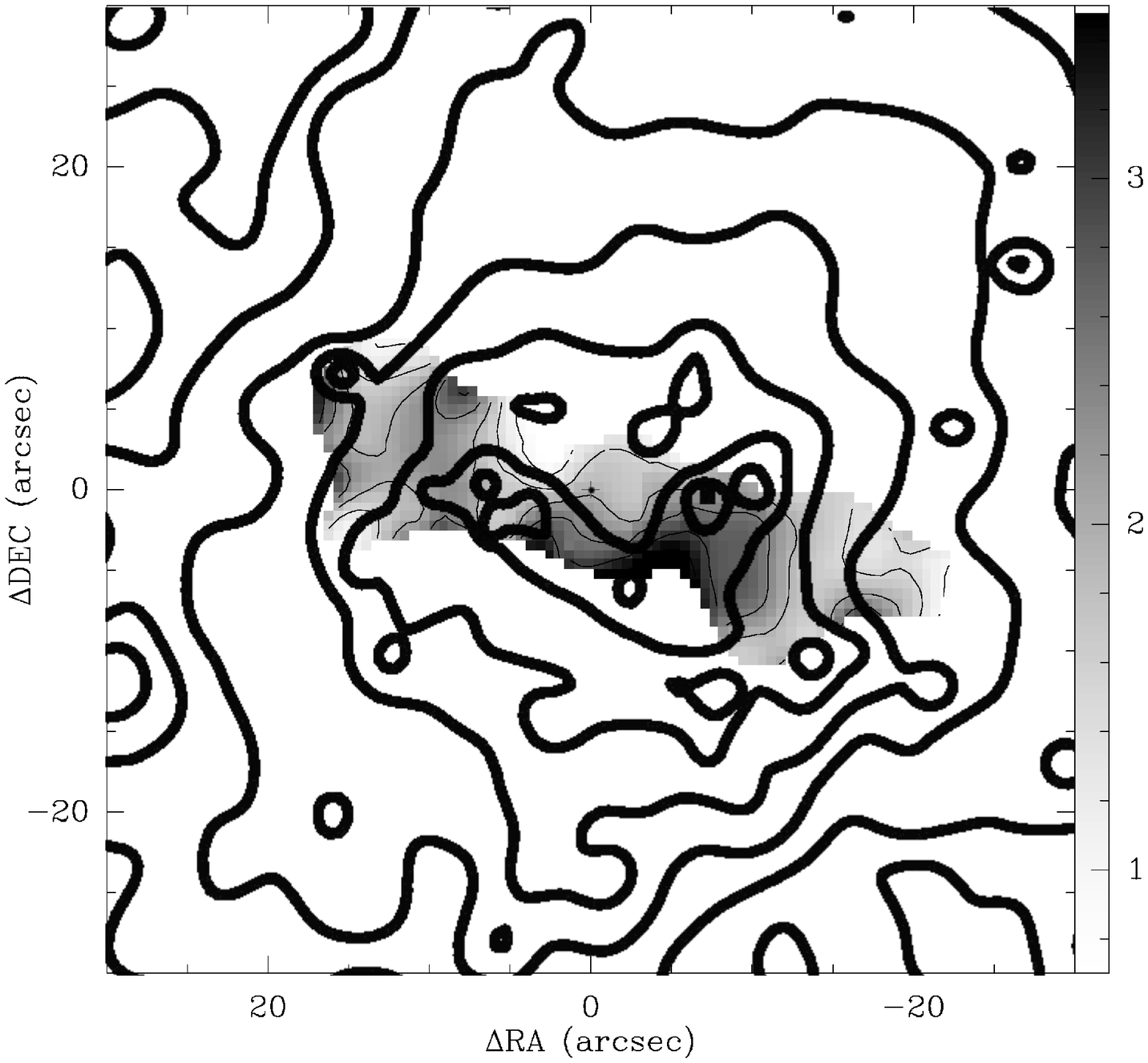}
\caption{Comparison between the spatial distribution of the diffuse hard X-ray emission 
(1.6-2.2 keV) with that of the HOC$^+$ emission ({\bf left panel}) and the
(H$^{13}$CO$^+$ 1$\rightarrow$0)/(HOC$^+$ 1$\rightarrow$0) brightness temperature ratio 
({\bf right panel}). Note the HOC$^+$ emisison is not spatially correlated
with the diffuse hard X-ray emission nor the (H$^{13}$CO$^+$ 1$\rightarrow$0)/(HOC$^+$ 1$\rightarrow$0)
brightness temperature ratio. 
(The X-ray image has been taken from Strickland \& Heckman 2007).}
\label{fig4}
\end{figure*}

\setlength\unitlength{1cm}
\begin{figure}
\vspace{13cm}
\includegraphics{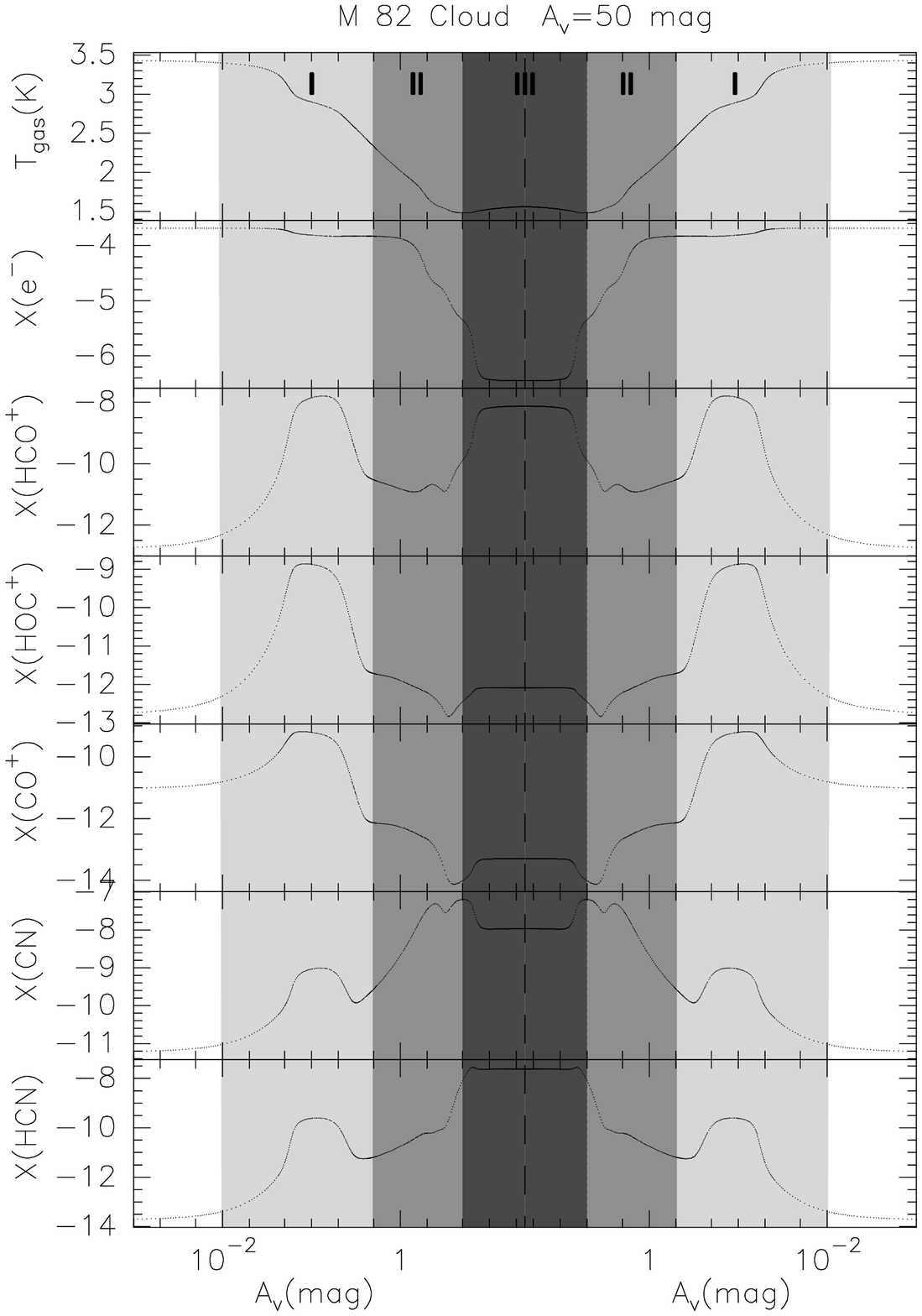}
\caption{Predictions for the gas temperature, electronic abundance and abundances of various species derived with our improved
PDR model.  The calculations have been carried out for a plane-parallel cloud of 50~mag, with a uniform density of
$n$~=~4~10$^5$~cm$^{-3}$, that is being illuminated by a UV field of  G$_0$~=~1.0 10$^4$ (in units of the Habing field) from the two sides. 
The cosmic rays flux is set to $\zeta$~=~4~10$^{-15}$~s$^{-1}$. We have shadowed the
three regions discussed in Section 4 and 5. }
\end{figure}
\vskip 0.5cm

\section{Discussion}
Following we discuss the two scenarios proposed to explain the 
large reactive ions abundances in M~82, UV-photon and X-ray 
dominated chemistry, on basis of our new observational results and
model calculations.

\subsection{Is X-ray chemistry at work in M~82?}

Spaans and Meijerink (2007) proposed that the diffuse X-ray emission at 0.7~keV could contribute
to produce the high abundances of reactive ions measured towards M~82. To have a deeper insight 
into the chemistry of these ions, we have compared our high-angular resolution HOC$^+$ image with 
the spatial distribution of the X-ray emission.

Strickland and Heckman (2007) reported the 
images of the continuum diffuse X-ray emission in the central 2~kpc~$\times$~2~kpc of 
M~82 for the energy bands between 0.3 to 7.0 KeV. The diffuse, both soft and hard, X-ray emission
arises from the inner $\sim$350~pc (20$"$) of the galaxy and extends along the
outflow lobes. The soft X-ray emission (E$<$1.1 keV) is
heavily absorbed by the material in the galaxy plane and peaks in the 
outflow lobes (at z$>$20$"$). In contrast, the less absorbed diffuse hard X-ray emission 
(E$<$1.1 keV) is intense in the galaxy disk and
extends only $\sim$8$"$ and $\sim$12$"$ above and below the galaxy plane.

Since soft X-rays are heavily extinguished towards the galaxy plane,
a detailed comparison between the spatial distribution of the soft X-ray flux and the HOC$^+$ 
emission is not possible. However, it is clear from the X-ray images that 
the soft X-rays are associated with the superwind itself and the superwind sources. The 
superwind is generated in the inner 350~pc of the galaxy where intense diffuse hard X-ray 
emission is detected.

In Fig. 5 we superpose the image of the diffuse hard X-rays emission
in the band 2.2-2.8 keV published by Strickland 
and Heckman (2007) on our HOC$^+$ and the [H$^{13}$CO$^+$]/[HOC$^+$] images.
Within the disk plane, the X-ray emission is concentrated in the region enclosed by the {\it inner ring} 
where only weak HOC$^+$ emission is found.
In fact the two most intense HOC$^+$ maxima are located just outside the X-rays emitting region.
Even towards the center of the galaxy, the most intense HOC$^+$ peak seems to avoid the peaks
of the X-ray emission.

There is no obvious relationship between the values of the
[H$^{13}$CO$^+$]/[HOC$^+$] ratio and the diffuse hard X-ray emission, either. 
Unlike the X-ray emission, 
the [H$^{13}$CO$^+$]/[HOC$^+$] ratio does not present any systematic trend with the distance from 
the galaxy center. Moreover, the  [H$^{13}$CO$^+$]/[HOC$^+$] ratio presents the highest value around 
SNR 41.95+57.5, the most intense X-ray source of the galaxy. 
Therefore, morphological arguments do not support the scenario of X-rays being the
driving agent of the reactive ions chemistry in M~82. 

An independent argument against the XDR scenario comes from the different spatial distribution
of the HOC$^+$ and SiO emissions in M~82. Usero et al. (2004) detected 
large SiO and HOC$^+$ abundances towards the circumnuclear region (CND) in NGC~1068. 
In this case, the large SiO and HOC$^+$ abundances were interpreted in terms of the 
interactions between the X-rays and, respectively, the dust and gas in the CND. 
Unlike NGC~1068, the SiO and the HOC$^+$ emissions arise in different regions
towards M~82. While the HOC$^+$ emission arises in the galaxy disk (this Paper), 
the SiO emission is tracing the gas outflowing from the galaxy plane  (Garc\'{\i}a-Burillo et al.
2001). This different spatial distribution suggest a different
origin for the HOC$^+$ and SiO emission in M~82 from that in NGC~1068.

\subsection{PDR model}
In this section, we explore the possibility of explaining the molecular abundances observed in
M~82 in terms of PDR chemistry. For this aim, we have used the updated version of the Meudon PDR code
(Le Petit et al. 2006).
The chemical network has been enlarged, relative to Paper III, to include the chemistry 
of HOC$^+$. Some reactions rates 
have also been updated according to the Ohio State University database (http://www.physics.ohio-state.edu/~eric/). 
In particular, 
the reaction rate of C$^+$ + OH $\rightarrow$ CO$^+$ + H has been increased to 2.9~10$^{-9}$~cm$^{-3}$~s$^{-1}$,
which translates into a larger CO$^+$ production and a better agreement between our predictions and the observational
results as we will discuss below. We have also included a new CO$^+$ formation reaction, 
O + CH$^+$ $\rightarrow$ CO$^+$ + H, but its contribution to the total CO$^+$ formation is very low ($<$3\%).

In Paper III, we fitted the [CO$^+$]/[HCO$^+$], [HCO$^+$]/[HOC$^+$] and [CN]/[HCN] ratios with a plane-parallel
model, assuming G$_0$=10$^4$ in units of the Habing field, n=4~10$^5$~cm$^{-3}$, a cosmic ray flux of 
4~10$^{-15}$~s$^{-1}$ and clouds with a total thickness of $\sim$~13~mag. The thickness of the clouds was estimated
as twice the thickness of the plane-parallel one side illuminated PDR that fits the
observations. Our model was quite succesful in predicting the  
[CO$^+$]/[HCO$^+$], [HCO$^+$]/[HOC$^+$] and [CN]/[HCN]
abundance ratios but could not account for the measured CO$^+$ column densities.  
We needed about $\sim$20--40 clouds along
the line of sight to explain the CN column densities, but then 
the predicted CO$^+$ column densities fell short by one order of magnitude. 
Conversely,
if we fitted the CO$^+$ column densities, the predicted CN column densities were far in
excess. This option looked less plausible because of 
the large number of clouds required to fit the CO$^+$ column densities ($>$100).

With our new chemical network, we obtain CO$^+$ column densities a factor of $\sim$5 larger 
than before (see Table 5). To improve our model, we have assumed , in addition,
a plane-parallel  cloud illuminated from the two sides. This is a more realistic view of 
the ISM in M~82. The parameters of our PDR model are shown in Table~4.

In Fig. 6 we show our model predictions for a plane-parallel cloud with a total visual extinction of 50~mag. 
The physical and chemical conditions are symmetric since the cloud is illuminated from the two sides by
the same UV field. We can distinguish three regions across the cloud according to their chemical properties: 
{\bf (I)} A$_v$=0.01--0.5~mag from the cloud surface. CO$^+$, HCO$^+$, HOC$^+$, CN and HCN present an abundance 
peak in this region and [CN]/[HCN]$\sim$20 . {\bf (II)} A$_v$=0.5--5~mag from the cloud surface. 
The abundances of CO$^+$, HCO$^+$, HOC$^+$ decline in this region. CN and HCN are, however, very
abundant and [CN]/[HCN]$\sim$100. {\bf (III)} A$_v$$>$5~mag from the cloud surface. 
The reactive ions CO$^+$ and HOC$^+$ have negligible abundances but
HCO$^+$, CN and HCN are abundant with [CN]/[HCN]$\sim$1. Note that the reactive 
ions CO$^+$ and HOC$^+$ only survive in region I. 
Therefore, reactive ions are excellent tracers  
of the number of clouds since their total column density remains
constant regardless of the thickness of the cloud for clouds with A$_v$$>$2~mag. 

In Tables 5, 6 and 7, we compare the predicted 
molecular column densities and abundance ratios with those observed towards M~82. 
In addition to CO$^+$, HOC$^+$, HCO$^+$, CN, HCN and HCO, we include the H$_3$O$^+$ column density derived
by van der Tak et al. (2008) and the upper limit to the H$_2$O column density obtained by 
Wilson et al. (2007). Thus, the most relevant molecules for the PDR/XDR chemistry are considered.

In Table 6 we show a comparison between our one-type-cloud model
and the column densities measured towards the E. position in M~82 (Paper I, II, III;
van der Tak et al. 2008). This is the position towards which the emission of most molecules
peak and column densities are better determined. The number of clouds along the line 
of sight have been calculated using the HOC$^+$ column densities. 
To match the observed
HOC$^+$ column density, we use 44 clouds of Av=5. This choice provides
a good fit (within a factor of 3) to many of the other observed column 
densities (see Table 7). However,
the model falls short by 2 orders  of magnitude to predict the HCN column densities. Related to it, 
the [CN]/[HCN] ratio is $\sim$380, far in excess the observed value. The model does not reproduce
the HCO and H$_2$O column densities either.
Moving to $\sim$10~mag clouds, we need 40 clouds along the line of sight to fit the HOC$^+$ column density. 
In this case, we improve our fit to the HCN column density and the [CN]/[HCN] ratio, 
but the predicted CN and H$_3$O$^+$ column densities are then too large. This is the same 
problem we found in Paper III. We cannot fit the CN, HCN, and reactive ions column densities 
simultaneously. The new reaction rates improve the fit but not enough. 
The fit would become worse if we go to larger clouds. The [CN]/[HCN] ratio would decrease and get 
closer to the observed value but the predicted CN, HCN, HCO$^+$ and H$_3$O$^+$ column densities 
would increase and become much larger than the observed values.

In order to look for a better solution, we have tried a two-cloud-type model. We obtain a good 
agreement with the observations by assuming that we have $\sim$44 clouds, 98.5\% of which have a 
thickness $\sim$5~mag  but a few percentage, 1.5\%, have a thickness of 50~mag. 
Since the assumed cloud morphology (a column plane-parallel cloud) is not realistic,
the number of clouds has here only a statistically meaning. It means that the emission
can be explained with a cloud distribution such that most of the molecular gas
(87\%)is forming small clouds (represented by our plane-paralell Av=5~mag cloud) while
only a few percentage (13\%) is in larger molecular clouds (represented by our 
plane-paralell Av=50~mag cloud). Compared with the case of 100\% of small clouds,
the small percentage of large clouds inject the lacking HCN column density and push down the CN/HCN ratio 
to fit the observed value of $\sim$6 without changing that much the [CO$^+$]/[HCO$^+$] and 
[HCO$^+$]/[HOC$^+$] ratios. 
In addition, our predicted HCO column densities are much closer, although still below,
the observed value. Note that the number of large clouds is limited by the HCN and HCO$^+$
column densities. Both species are very abundant in region III of our pattern 50~mag cloud
and their column densities increase rapidly with the number of large clouds.
With the proposed mix, we fit all the observed column densities except those of
HCO and H$_2$O. 

Our model falls short by a factor of 10 to predict the observed HCO column densities. 
Schilke et al. (1992) proposed that
the photodissociation of the H$_2$CO molecules released from the grain mantles 
could increase the HCO abundance in the moderately illuminated regions of the PDR.
Our model does not include production of these species on
grains.

Regarding H$_2$O, the predicted column densitites are 2 orders of magnitude larger than the upper limit obtained
by Wilson et al. (2007). This discrepancy is very likely because we are using a gas-phase
model that neglects the depletion of water in dense clouds. 
It is also important to recall that this limit 
was obtained with a large beam (2.1$'$) and by observing the 
556.936 GHz transition of ortho-H$_2$O which is very likely optically thick. 
Therefore, the limit is quite uncertain. 

Summarizing, we conclude that the most relevant PDR molecules are well fitted
by assuming that most of the mass of the molecular 
gas (87\%) is forming small clouds with a total thickness of $\sim$5~mag. A small percentage, however, 
is encompassed in large molecular clouds, $\sim$50~mag, in which star formation might be proceeding.
This suggests that the mass distribution of the clouds in M~82 is very different from that in our Galaxy, and
most of the giant molecular clouds (GMCs) have already been destroyed by the recent starburst.
M~82 is, therefore, an old starburst where star formation has almost exhausted the molecular
gas reservoir.

\section{Conclusions}
In this paper, we present new JCMT observations of the HCO$^+$ and HOC$^+$ 3$\rightarrow$2 and
4$\rightarrow$3 lines and the PdB interferometric image of the HOC$^+$ 1$\rightarrow$0 line 
towards the nucleus of M~82.
Our results can be summarized as follows:

\begin{itemize}
\item We have carried out excitation calculations for HCO$^+$ and HOC$^+$ taking into account
our 30m and JCMT observations of the HCO$^+$ and HOC$^+$ J=3$\rightarrow$2 and J=4$\rightarrow$3 
lines.  Our results show that the two ions arise in dense gas (n$>$10$^4$~cm$^{-3}$). Therefore,
HOC$^+$ is likely tracing the partly ionized and moderately dense envelopes of
molecular clumps, rather than in the intercloud medium or wind phase.

\item We present a high angular resolution (4$"$) image of HOC$^+$ 1$\rightarrow$0 in M~82,
the first ever obtained in Galactic and extragalactic research. The comparison between
the diffuse X-ray emission and our HOC$^+$ image discards X-rays as the major
driver of the high abundances of reactive ions in M~82.
UV photons from massive stars seem, therefore, the major drivers of the molecular chemistry in M~82.

\item 
Our PDR chemical model shows that most molecular abundances measured towards M~82 
(HCO$^+$, HOC$^+$, CO$^+$, CN, HCN, H$_3$O$^+$) are well accounted in the scenario 
of two-types (different thickness, A$_v$=5~mag and A$_v$=50~mag) of dense 
clouds ($>$10$^4$ cm$^{-3}$) bathed
by an intense interstellar UV field (G$_0$=10$^4$). 

\item Furthermore, our model shows that most of the molecular gas, 
87 \% in mass, is forming small  clouds (A$_v$$\sim$5~mag) while only 13\%
is in large molecular clouds (A$_v$$\sim$50~mag). This suggests that 
only a small percentage ($<$13\%) of the molecular gas in M~82 is 
regions where massive star formation is still proceeding. 
\end{itemize}

\begin{table}
\caption{PDR model }
\begin{tabular}{ll} \hline
G$_0$    (Habing)    & 1~10$^4$  \\
n        (cm$^{-3}$)  & 4~10$^5$   \\
$\zeta$  (s$^{-1}$)   &  4 10$^{-15}$\\
He      &  0.1 \\
C       &  1.32 10$^{-4}$ \\ 
O       &  3.19 10$^{-4}$ \\   
N       &  7.50 10$^{-5}$ \\   
S       &  1.86 10$^{-5}$ \\
Fe      &  1.50 10$^{-8}$ \\
\hline
\end{tabular}
\end{table}

\begin{table}
\caption{Model Results (I) }
\begin{tabular}{ccccc} \hline
\multicolumn{1}{c}{A$_v$} &   \multicolumn{1}{c}{$\frac{CN}{HCN}$} & 
\multicolumn{1}{c}{$\frac{HCO^+}{HOC^+}$} &  \multicolumn{1}{c}{$\frac{CO^+}{HCO^+}$} & 
\multicolumn{1}{c}{$\frac{CN}{CO^+}$} \\
\multicolumn{1}{c}{(mag)}       &   & 
                                &   &   \\
\hline\hline
M~82 &  6   & 44  & 0.04  &  420 \\
2    &  18  & 12  & 0.04  &  7  \\
5    &  380 & 12  & 0.04  &  479 \\
10   &  38  & 13  & 0.04  &  2129 \\
30   &  1.3 & 305 & 0.002 &  4193 \\
50   &  0.9 & 671  &  7 10$^{-4}$  &   5397 \\
100  &  0.6 & 1500 &  3 10$^{-4}$  &   8519 \\
\hline
\hline
\end{tabular}
\end{table}

\begin{table*}
\caption{Model Results (II) }
\begin{tabular}{ccccccccc} \hline
\multicolumn{1}{c}{A$_v$} & \multicolumn{1}{c}{N(CO$^+$)} & \multicolumn{1}{c}{N(HOC$^+$)} &
\multicolumn{1}{c}{N(HCO$^+$)} & \multicolumn{1}{c}{N(CN)} & \multicolumn{1}{c}{N(HCN)} &
\multicolumn{1}{c}{N(HCO)} & \multicolumn{1}{c}{N(H$_3$O$^+$)} & 
\multicolumn{1}{c}{N(H$_2$O)} \\
\multicolumn{1}{c}{(mag)}       &  \multicolumn{1}{c}{(cm$^{-2}$)} & 
\multicolumn{1}{c}{(cm$^{-2}$)} &  \multicolumn{1}{c}{(cm$^{-2}$)} & 
\multicolumn{1}{c}{(cm$^{-2}$)} &   \multicolumn{1}{c}{(cm$^{-2}$)} & 
\multicolumn{1}{c}{(cm$^{-2}$)} & \multicolumn{1}{c}{(cm$^{-2}$)} &
\multicolumn{1}{c}{(cm$^{-2}$)} \\
\hline\hline
M~82 & 1.5 10$^{13}$ & 2.5 10$^{13}$ & 1.1 10$^{15}$ & 6.3 10$^{15}$ &  1.1 10$^{15}$ & 4.4 10$^{12}$ & 1.1 10$^{14}$ & $<$2 10$^{14}$ \\
2    & 2.4 10$^{11}$ & 4.5 10$^{11}$ & 5.5 10$^{12}$ & 1.7 10$^{12}$ &  9.4 10$^{10}$ & 2.1 10$^{9}$ & 4.9 10$^{12}$ & 4.7 10$^{13}$ \\
5    & 2.9 10$^{11}$ & 5.7 10$^{11}$ & 7.2 10$^{12}$ & 1.4 10$^{14}$ &  3.6 10$^{11}$ & 3.7 10$^{9}$ & 6.0 10$^{12}$ & 6.6 10$^{13}$ \\
10   & 3.1 10$^{11}$ & 6.2 10$^{11}$ & 8.2 10$^{12}$ & 6.6 10$^{14}$ &  1.7 10$^{13}$ & 4.4 10$^{9}$ & 6.8 10$^{12}$ & 1.1 10$^{15}$ \\
30   & 3.2 10$^{11}$ & 6.7 10$^{11}$ & 2.0 10$^{14}$ & 1.4 10$^{15}$ &  1.0 10$^{15}$ & 2.1 10$^{12}$ & 6.9 10$^{13}$ & 4.8 10$^{16}$ \\
50   & 3.3 10$^{11}$ & 7.1 10$^{11}$ & 4.8 10$^{14}$ & 1.8 10$^{15}$ &  2.0 10$^{15}$ & 7.8 10$^{12}$ & 1.5 10$^{14}$ & 1.0 10$^{17}$ \\
100  & 3.4 10$^{11}$ & 8.0 10$^{11}$ & 1.2 10$^{15}$ & 2.9 10$^{15}$ &  4.6 10$^{15}$ & 2.2 10$^{13}$ & 3.7 10$^{14}$ & 2.4 10$^{17}$ \\
\hline
\hline
\end{tabular}
\end{table*}

\begin{table*}
\caption{Model Results (III) }
\begin{tabular}{lllll} \hline
\multicolumn{1}{c}{} & 
\multicolumn{1}{c}{M82}  & 
\multicolumn{1}{c}{100\%A$_v$=5~mag}  & \multicolumn{1}{c}{100\% A$_v$=10~mag} & 
 \multicolumn{1}{c}{98.5\% A$_v$=5~mag}  \\ 
\multicolumn{1}{c}{} & \multicolumn{1}{c}{}  & \multicolumn{1}{c}{} & \multicolumn{1}{c}{}  &
 \multicolumn{1}{c}{1.5\% A$_v$=50~mag}  \\ 
\hline\hline
            & M~82                        &   No. clouds=44      &      No. clouds=40        &   No clouds=44         \\
N(CO$^+$)   &  1.5 10$^{13}$              &   1.3 10$^{13}$      &      1.2 10$^{13}$        &   1.3 10$^{13}$        \\
{\it N(HOC$^+$)} &  {\it 2.5 10$^{13}$}   & {\it 2.5 10$^{13}$}  &    {\it 2.5 10$^{13}$}    &  {\it 2.5 10$^{13}$ }\\
N(HCO$^+$)  &  1.1 10$^{15}$              &  3.2 10$^{14}$       &  {\bf 3.3 10$^{14}$}$^{**}$  &  6.3 10$^{14}$       \\
N(CN)       &  6.3 10$^{15}$              &  6.1 10$^{15}$       &   {\bf 2.6 10$^{16}$}        &  7.2 10$^{15}$       \\
N(HCN)      &  1.1 10$^{15}$              & {\bf 1.6 10$^{13}$}  &       6.8 10$^{14}$          &  1.4 10$^{15}$       \\
H(HCO)      &  4.9 10$^{13}$$^*$          & {\bf 1.6 10$^{11}$}  &   {\bf 1.7 10$^{11}$}        & {\bf 5.3 10$^{12}$}   \\
N(H$_3$O$^+$) & 1.1 10$^{14}$             &   2.6 10$^{14}$      &  {\bf 2.7 10$^{14}$}         &  3.6 10$^{14}$  \\
N(H$_2$O)     & $<$2 10$^{14}$            &  {\bf 2.8 10$^{15}$} &  {\bf 4.2 10$^{16}$}  &         {\bf 6.9 10$^{16}$}  \\
N(CO$^+$)/N(HCO$^+$)  & 0.04    &    0.04               &  0.04                &    0.02    \\
N(HCO$^+$)/N(HOC$^+$) & 44      &    12                 &  13                  &    25      \\
N(CN)/N(HCN)          & 6       &    380                 &  38                  &    5     \\
\hline
\hline
\end{tabular}

\noindent
$^*$ Average value in our HCO interferometric image (Garc\'{\i}a-Burillo et al. 2002)

\noindent
$^{**}$ We mark in bold face those predictions that differ by more than a factor of 3 from the observed values
\end{table*}
\vskip 0.5cm

\begin{acknowledgements}
We thank the IRAM and JCMT staff for their help and support during the observations and data reduction. 
AU has been supported through a Post Doctoral Research Assistantship from 
the UK Science \& Technology Facilities Council.
We are also grateful to C. Joblin, O. Bern\'e and J.R. Goicoechea 
for fruitful discussions on the PDR chemistry.
\end{acknowledgements}

\appendix

\section{Rotational rate coefficients}

For the rotational excitation of HCO$^+$ by hydrogen molecules, we
employed the rate coefficients computed by Flower (1999). These rates
are available for transitions between levels with $J\leq 20$ and
temperatures in the range 10 $\leq T \leq$ 400~K. For HOC$^+$,
de-excitation rates were assumed identical to those of HCO$^+$ and the
detailed balance principle was applied to derive the HOC$^+$
excitation rates. Note that these rates correspond to H$_2$ in its
ground para level ($J=0$) and that rates for H$_2$ in excited
rotational levels are not available.

Rate coefficients for electron-impact rotational excitation of HCO$^+$
and HOC$^+$ were computed using the $R$-matrix method combined with
the Adiabatic-Nuclei-rotation (ANR) approximation. The {\it ab initio}
$R$-matrix model of Faure \& Tennyson (2001) for HCO$^+$ was employed
here for all calculations on both molecular ions. The (theoretical)
equilibrium geometry of HOC$^+$ was taken from Ma et al. (1992). A
ground state dipole moment of 2.797~D was obtained, in excellent
agreement with the value of 2.8~D computed by Defrees et
al. (1982). The two lowest target states of symmetry $^1\Sigma^+$ and
$^3\Pi$ were included in the close-coupling calculation. Fixed-nuclei
$T$-matrices were obtained for total symmetry $^2\Sigma^+$, $^2\Pi$
and $^2\Delta$, as in the case of HCO$^+$. Full details about the
$R$-matrix model will be published elsewhere.

The major difference with the calculations of Faure \& Tennyson (2001)
concerns the application of the ANR approximation. This latter has
been recently shown to be valid down to threshold for molecular ions,
provided a simple ``Heaviside correction'' is applied to excitation
cross sections. Full details can be found in Faure et al. (2006). Note
that closed-channel effects are neglected here as these are expected
to be small for strongly polar ions (see Faure et al. 2006). 
Furthermore, in contrast to the calculations of Faure \&
Tennyson (2001) restricted to the lowest 3 rotational levels, we have
considered here rotational transitions between levels with $J\leq 20$
(i.e. involving rotational energies up to $\sim$900~K). Transitions
were however restricted to $\Delta J\leq 8$ owing to the limited
number of partial waves included in the $T$-matrices ($l\leq$
4). Finally, de-excitation rates have been fitted over the analytic
form, Eq.~(2), of Faure et al. (2004) for temperatures in the range 5
$\leq T \leq$ 1000~K. Fitting coefficients for both HCO$^+$ and
HOC$^+$ will be published in a forthcoming publication (Faure \&
Tennyson, in preparation). Note, finally, that the present HCO$^+$
rates have also been employed recently by Jim{\'e}nez-Serra et
al. (2006) to explain the overexcitation of HCO$^+$ in the shock
precursor component of the young L1448-mm outflow.

\end{document}